\documentclass[twocolumn, superscriptaddress, prb, showpacs]{revtex4}

\usepackage{graphicx}
\usepackage{amsmath}

\newcommand{\sfA}{{\sffamily A}}
\newcommand{\sfB}{{\sffamily B}}
\newcommand{\sfC}{{\sffamily C}}

\begin{document}

\title{Fcc breathing instability in BaBiO$_3$ from first principles}
\author{T. Thonhauser}
\affiliation{Department of Physics and Astronomy, Rutgers, The State
University of New Jersey, Piscataway, New Jersey 08854, USA.}
\author{K. M. Rabe}
\affiliation{Department of Physics and Astronomy, Rutgers, The State
University of New Jersey, Piscataway, New Jersey 08854, USA.}
\date{\today}

\begin{abstract}
We present first-principles density-functional calculations using the
local density approximation to investigate the structural instability
of cubic perovskite BaBiO$_3$. This material might exhibit charge
disproportionation and some evidence thereof has been linked to the
appearance of an additional, fourth peak in the experimental IR
spectrum. However, our results suggest that the origin of this
additional peak can be understood within the picture of a simple
structural instability. While the true instability consists of an
oxygen-octahedra breathing distortion and a small octahedra rotation,
we find that the breathing alone in a fcc-type cell doubling is
sufficient to explain the fourth peak in the IR spectrum. Our results
show that the oscillator strength of this particular mode is of the
same order of magnitude as the other three modes, in agreement with
experiment.
\end{abstract}

\pacs{77.80.-e, 71.20.-b, 63.20.-e, 78.30.-j}
\maketitle

Over the last few decades, BaBiO$_3$ has been a subject of continuing
interest for several reasons. First, this material is an end member of
the series (Ba,K)(Pb,Bi)O$_3$, which includes one of the few
high-temperature superconductors that do not contain copper. Second,
the formal valence of Bi for the cubic perovskite BaBiO$_3$ is $4+$,
suggesting that the Bi should exhibit charge disproportionation
$2\,\text{Bi}^{4+}\to \text{Bi}^{3+}+\text{Bi}^{5+}$ leading to a
charge-ordered state. A better understanding of such ordered states
could lead to the design of new charge-ordered ferroelectrics, a class
of materials that has recently attracted considerable
attention.\cite{khomskii04} Experimental evidence for static charge
disproportionation in BaBiO$_3$ has been linked to the interpretation
of an additional, fourth strong peak in its IR
spectrum,\cite{uchida85,gervais95,nishio03,ahmad04} while group theory
predicts only three peaks for the cubic perovskite structure.

The goal of this paper is to show that not only charge
disproportionation, but also a much simpler mechanism---i.e.\ a
structural breathing instability---can account for all four peaks in
the experimental IR spectrum of BaBiO$_3$. To this end, we first
perform density-functional theory (DFT) calculations to investigate the
structural instability of cubic perovskite BaBiO$_3$. Thereafter, we
calculate the IR active phonons and the Born effective charges, which
in turn allows us to determine the IR oscillator strength of all four
modes mentioned above. We would also like to point out that for our
calculations we assumed a simple fcc cell-doubling breathing
distortion, which is sufficient to warrant non-zero oscillator
strengths for all four modes. The full structural instability, however,
consists not only of a breathing of the oxygen octahedra, but also of a
small nearly-rigid rotation thereof.\cite{cox76, liechtenstein91,
kunc91}

Our first-principles density-functional-theory calculations were
performed using the VASP package\cite{vasp1} and PAW
potentials\cite{vasp2} with a 520~eV kinetic-energy cutoff. Note that
we included the Ba 5$s^2$ and Bi 5$d^{10}$ states as valence states.
The exchange-correlation energy functional was evaluated within the
local density approximation (LDA) as parameterized by Perdew and
Zunger.\cite{Perdew81} Brillouin-zone integrals were approximated with
the Monkhorst--Pack scheme on a 14$\times$14$\times$14 mesh of special
k-points.\cite{monkhorst76} Spin-orbit effects were not taken into
account, since previous calculations showed that their effect on the
band structure around the Fermi energy is small.\cite{mattheiss83}

\begin{figure}
\begin{center}
\includegraphics[width=3.8cm]{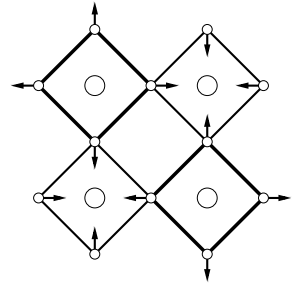}
\end{center}
\caption{\label{fig:distortion}Distortion of the oxygen octahedra.
Small circles denote oxygen atoms and big circles denote bismuth atoms,
respectively. In three dimensions, this breathing distortion produces a
checkerboard pattern of adjacent smaller and larger octahedra, doubling
the primitive simple cubic unit cell to fcc.}
\end{figure}

The high-temperature phase of BaBiO$_3$ above 750~K has the simple
cubic perovskite structure with five atoms per unit cell.\cite{cox76}
The simple cubic lattice constant of 4.328~\AA\ for our calculations
was derived from reported experimental values. In this structure,
BaBiO$_3$ has an odd number of electrons per unit cell and therefore is
metallic. However, lower-temperature experiments and calculations show
that the unit cell for the low temperature phase is
monoclinic,\cite{cox76, liechtenstein91, kunc91} and that the material
is a semiconductor with an estimated gap of 0.2~eV.\cite{sleight75} The
monoclinic phase results from a distortion of the cubic cell consisting
mainly of a breathing of the oxygen octahedra and a small nearly-rigid
rotation thereof. As mentioned above, for our investigation we limit
ourselves to the pure breathing of the octahedra, which is enough to
explain the appearance of all four peaks in the IR spectrum. In the
breathing distortion, the oxygen atoms can move along the Bi--O bond to
produce two inequivalent Bi sites with different Bi--O bond lengths,
without breaking any point symmetries. This results in a doubling of
the unit cell to ten atoms, following the three dimensional
checkerboard pattern sketched in Fig.~\ref{fig:distortion}. The
resulting fcc lattice has a conventional lattice constant of 8.655~\AA.
This distortion has previously been proposed by Uchida and
coworkers.\cite{uchida85}

\begin{figure}
\begin{center}
\includegraphics[width=\columnwidth]{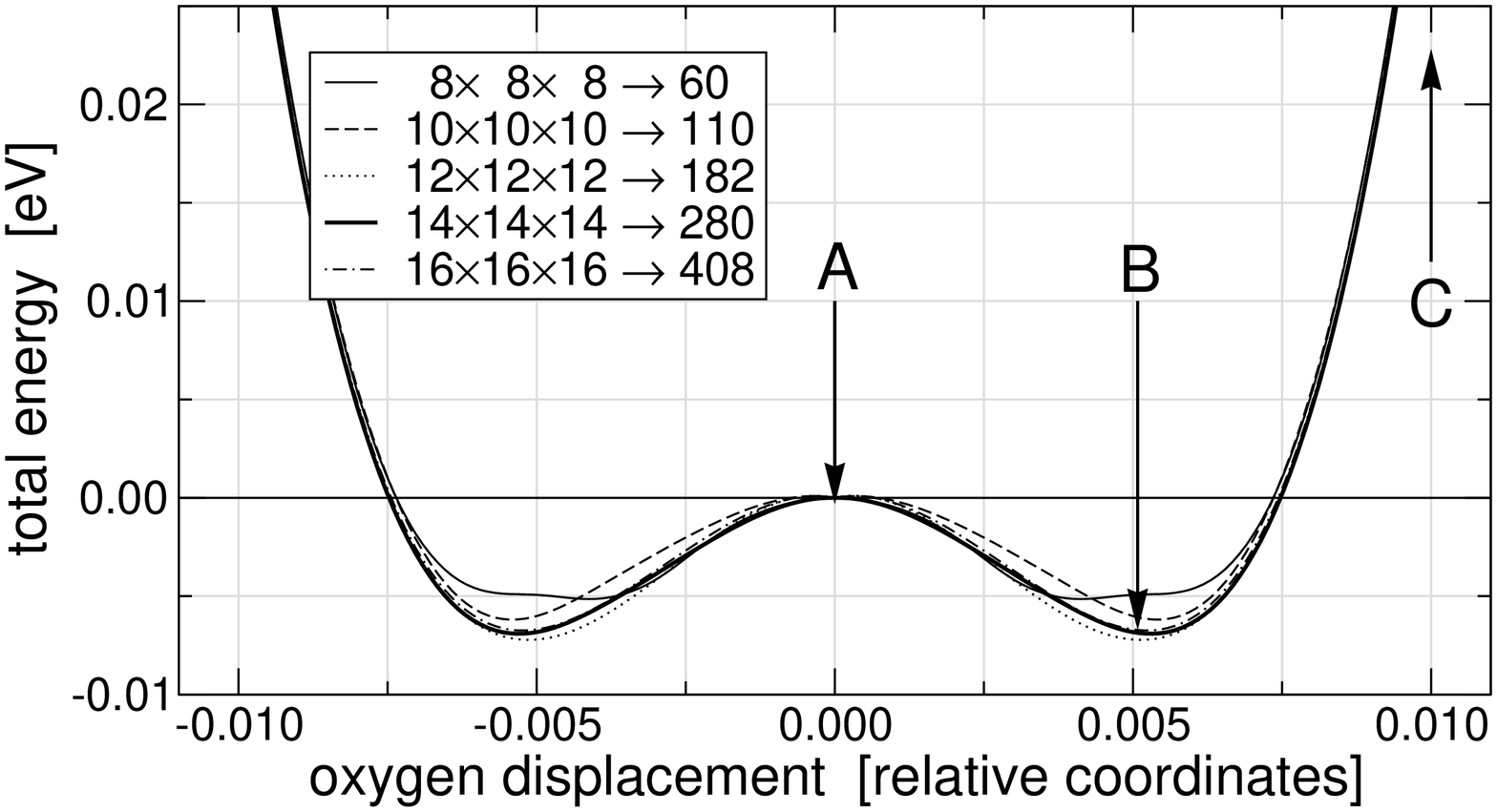}
\end{center}
\caption{\label{fig:double_well}K-mesh dependence of the total energy
of  fcc BaBiO$_3$ as a function of oxygen displacement from the high
symmetry position of 1/4 (point \sfA) given in relative Cartesian
coordinates. Results are plotted for meshes from $8\times 8\times 8$
(i.e. 60 k-points in the IBZ) up to $16\times 16\times 16$ (i.e. 408
k-points in the IBZ). Convergence is reached with a $14\times 14\times
14$ mesh, which produces results indistinguishable from any denser
mesh. The minima of the double-well potential are at $\pm$0.005 (\sfB),
resulting in octahedra of two different sizes.}
\end{figure}

Figure~\ref{fig:double_well} depicts the resulting double-well
potential when the oxygen atoms are moved along the Bi--O bonds in the
vicinity of the high symmetry oxygen position of 1/4, marked as \sfA.
The different curves correspond to different dense k-meshes. Our
results show that a $8\times 8\times 8$ mesh, i.e. 60 k-points in the
irreducible part of the Brillouin zone (IBZ), which is usually
sufficient for calculations on similar perovskites, does not provide
converged results. A very dense mesh of $14\times 14\times 14$ k-points
(i.e. 280 k-points in IBZ) is necessary to describe this sensitive
breathing distortion correctly. This is partly due to the fact
that---as we will see below---the system is close to the
metal/semiconductor transition. The double well has minima of depth
6.7~meV per 10-atom unit-cell at a displacement of $\pm$0.00507 (\sfB),
corresponding to the symmetry-equivalent choices of breathing a given
octahedron either in or out. The shape of our double-well potential
agrees well with LMTO calculations for breathing where a rotation of
13$^\circ$ has been frozen in.\cite{meregalli98} Other calculations
find for the double-well without rotation a depth of 20~meV at a larger
displacement of 0.07~\AA\ (LMTO)\cite{kunc91} and a depth of 50~meV at
0.06~\AA\ (LAPW).\cite{blaha90} Lichtenstein and coworkers, including
combined tilting and breathing, find a minimum of 14~meV at 0.055~\AA\
with 8.5$^\circ$ rotation.\cite{liechtenstein91} Note that the
experimental values at 150~K are a rotation of 11.2$^\circ$ and a
displacement of 0.085~\AA.

Henceforth, we will abbreviate in this paper our calculated optimal
oxygen displacement as $\pm$0.00507$\approx\pm$0.005, keeping in mind
that all calculations were actually done at the exact minimum position.
At the minima, the oxygen atoms are displaced 0.04\AA. As a result, the
simple cubic Bi--O bond length of 2.164~\AA\ splits into 2.120~\AA\ and
2.208~\AA\ for Bi(1)--O and Bi(2)--O. We also studied the shape of the
double-well potential as a function of pressure. It turns out that
pressure decreases the depth of the double well and moves its minimum
closer to the high-symmetry position, thereby stabilizing the
structure.

\begin{figure}
\begin{center}
\includegraphics[width=7.5cm]{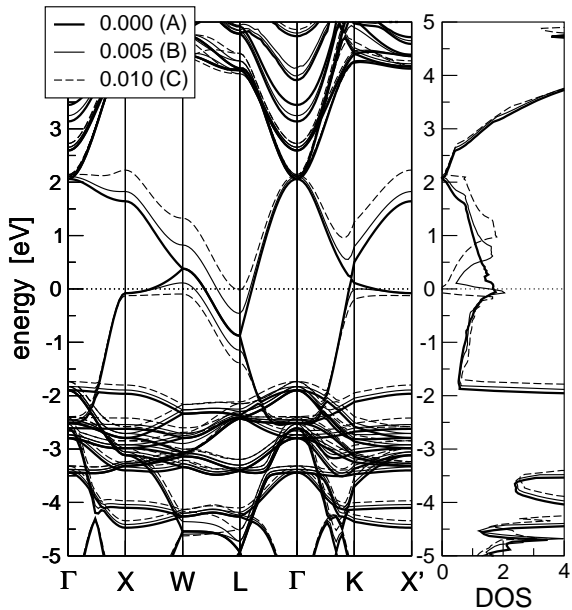}
\vspace{-4mm}
\end{center}
\caption{\label{fig:bs}Band structure and DOS of BaBiO$_3$ for the 
oxygen displacements \sfA, \sfB, and \sfC\ of
Fig.~\ref{fig:double_well}. Calculation \sfA\ corresponds to the simple
cubic perovskite cell and reflects the metallic character expected. The
oxygen displacement of 0.005 (\sfB) splits the degenerate band along
$W$--$L$ and decreases the DOS around the Fermi energy. The results are
plotted with respect to the Fermi energy and the DOS is given in
states/eV.}
\end{figure}

In order to investigate the effect of the breathing upon the gap, we
calculated the band structure and DOS for different oxygen
displacements as depicted in Fig.~\ref{fig:bs}.  It can be seen that at
the minima of the double-well (\sfB) a small band overlap along
$W$--$L$ is present, rather than a gap. This finding is in agreement
with previous calculations\cite{mattheiss83} and the discrepancy is
consistent with the well-known tendency of LDA to underestimate the
gap. For the larger displacement according to point \sfC, the gap is
clearly open. It can also be seen that the degenerate band along
$W$--$L$ in the simple cubic structure splits as the oxygens move. The
splitting is almost constant along the band and of the order of 1~eV.
Calculations by Mattheiss and Hamann, including the breathing and
tilting of the octahedra, show that this splitting is primarily due to
the breathing,\cite{mattheiss83} and is therefore captured correctly
within our approach. In addition, the bands at $K$ repel each other.
Both effects together provide a mechanism to open the gap in agreement
with the observed semiconducting character of
BaBiO$_3$.\cite{sleight75}

We next used a frozen-phonon approach to calculate the zone-center
phonon frequencies and eigenvectors for the minimum-energy fcc
structure with oxygen displacements of $\pm$0.005, corresponding to
point \sfB\ in Fig.~\ref{fig:double_well}. The optical modes in this
case are
\begin{equation}
\Gamma^{\text{opt}}=A_{1g}+E_g+T_{2u}+2T_{2g}+4T_{1u}+T_{1g}\;,
\end{equation}
four of which are IR active ($T_{1u}$). The symmetry adapted
calculation gives for these four modes the frequencies $\omega_1=444$
(438), $\omega_2=198$ (231), $\omega_3=157$ (136), and $\omega_4=32$
(98)~cm$^{-1}$, where the experimental values in parenthesis are taken
from Ref.~\onlinecite{nishio03}. Considering that we neglected the
rotational distortion, the agreement of the high-frequency modes is
reasonable, while the low-frequency mode at 31~cm$^{-1}$ differs
considerably from the experimental value.

Next, we calculated the Born effective charge tensors
$Z^*_{k,\alpha\beta}$ using a Berry-phase scheme.\cite{berry1,berry2}
This calculation could not be carried out directly for the
minimum-energy fcc structure, as it  is metallic due to a small band
overlap, as described above.  We computed the Born effective charges
for the semiconducting structure obtained by displacing the oxygens by
an additional 0.03\AA\ to the experimental displacement of 0.08\AA,
since at this point the gap has just opened. This procedure is
justified by the fact that the phonon corresponding to the oxygen
displacement is non-polar and the Born effective charges are expected
to be smooth near the metal/semiconductor transition. Indeed, we
investigated the dependence of the effective charges upon the oxygen
displacement by performing calculations for a larger displacement
corresponding to 0.010 (point \sfC\ in Fig.~\ref{fig:double_well}). Our
results show that the effective charges for the displacement 0.009 and
0.010 differ only by approximately 1\%. All $Z^*_{k,\alpha\beta}$ are
found to be diagonal and the results are collected in 
Table~\ref{tab:born}. The values for the effective charges are
reminiscent of similar perovskites, such as BaTiO$_3$.\cite{rabe99} The
splitting of $Z^*$ for Bi(1) and Bi(2) is quite striking, however, note
that this ``Born effective charge disproportionation" is not
necessarily the result of static charge transfer.

\begin{table}
\caption{\label{tab:born}Born effective charges $Z^*_{k,\alpha\beta}$ 
for atom $k$ in units of $|e|$ for fcc BaBiO$_3$. All 
$Z^*_{k,\alpha\beta}$ are diagonal. Bi(1) and Bi(2) are located in the
center of a larger and smaller octahedra, respectively. For the 
O-atoms we refer to $Z^*_{||}$ and $Z^*_{\bot}$, denoting the
components parallel and perpendicular to the Bi--O bond.}
\begin{tabular*}{\columnwidth}{l@{\extracolsep\fill}cccc}\hline\hline
Atom     &   $Z^*_{k,11}$, $Z^*_{k,22}$, $Z^*_{k,33}$
&Atom    &   $Z^*_{||}$ &  $Z^*_{\bot}$\\\hline
Bi(1)        &  $+4.78$ & O(1--6) & $-4.55$   & $-1.85$\\
Bi(2)        &  $+6.22$ &         &           & \\
Ba(1), Ba(2) &  $+2.75$ &         &           & \\\hline\hline
\end{tabular*}
\end{table}

For the simple cubic perovskite structure group theory predicts three
3-fold degenerate IR active phonons.\cite{uchida85}  When the unit cell
is doubled as described above, the zone-boundary $R$-point phonons are
folded back to the zone center. This leads to a mixing of one of the
folded-back phonons (with alternating Bi displacements relative to
their surrounding oxygen octahedra) and the IR active zone-center
phonons. In the five-atom cell this phonon has no oscillator strength
since the contributions of neighboring Bi cancel. In the doubled cell,
the octahedra of different sizes create different environments for the
bismuth atoms sitting in their centers and the contributions no longer
cancel. As a result, the oscillator strength of this particular mode
becomes non-zero, which explains the additional, fourth peak in the
experimental IR spectrum. Note that materials such as BaTiO$_3$ or
SrTiO$_4$, that have a single valence for the Ti ion (4+), only show
three strong peaks in the IR spectrum---as expected from group theory.

As the last step, we calculate the oscillator strengths of the IR
active modes. The oscillator strength $S_i$ of mode $i$ can be
calculated as
\begin{equation}
S_i =\sum_\alpha\bigg[\sum_{\beta,k}\frac{Z^*_{k,\alpha\beta}\;
   \xi_\beta(k,i)}{\sqrt{m_k}}\bigg]^2\;,
\end{equation}
where $\xi(k,i)$ is the displacement of atom $k$ according to the 
eigenvector of the $i$th phonon mode, $m_k$ is the mass of atom $k$,
and  $Z^*_{k,\alpha\beta}$ are the Born effective charges. Our results
for the oscillator strengths are $S_1=2.097$, $S_2=0.457$, $S_3=0.661$,
and $S_4=0.617$. The three modes $S_i$ ($i=1,2,4$) have been associated
with the simple cubic perovskite structure and can be identified in the
experimental spectrum.\cite{uchida85,gervais95,nishio03,ahmad04} The
oscillator strength of the fourth mode $S_3=0.661$ is of the same order
of magnitude as the ones associated with the simple cubic perovskite
structure. Indeed, as a result, a fourth peak is clearly visible in
experiments. The fact that our calculations predict an oscillator
strength for this mode comparable to the other modes goes beyond pure
group theory, which cannot make any predictions about the oscillator
strength. As our main result, we conclude that DFT calculations for a
pure fcc-type breathing can explain the existence of all four peaks in
the IR spectrum of BaBiO$_3$.

Further analysis of $S_3$ reveals that its value is almost independent
of the  Born-effective-charge splitting between the two inequivalent Bi
atoms. If we remove the splitting between $Z^*_{\text{Bi(1)}}=4.78$ and
$Z^*_{\text{Bi(2)}}=6.22$ by hand and replace these values with the
average of 5.5, the oscillator strength changes only by a few percent.
It follows that the oscillator strength is mainly determined by the
eigenvector of the corresponding phonon mode. Furthermore, we find that
about half of the oscillator strength is a result of the asymmetry in
displacement of the Bi(1) and Bi(2) atoms. The rest is mostly due to
the displacement of the four oxygen atoms in the plane perpendicular to
the Bi movement.

To conclude, we have performed first-principles DFT calculations to
investigate the electronic structure and structural instability of
cubic perovskite BaBiO$_3$. We investigate the effect of a fcc-like
unit-cell doubling produced by a breathing distortion of the oxygen
octahedra. The oxygen atoms are thereby displaced 0.04\AA\ along the
Bi--O bond, which generates smaller and larger octahedra throughout the
crystal. The distortion is very small, and yet we find that properties
such as the band gap or the Born effective charges are very sensitive
to it. This cell-doubling distortion turns-on the oscillator strength
of an additional fourth mode, observed as an additional strong peak in
experiment. In particular, our calculations show that the oscillator
strength of this certain mode is of the same order of magnitude as the
strengths of the remaining three modes.  In future work, we would like
to improve our results by including not only the breathing distortion,
but also the rotation of the octahedra. Furthermore, although BaBiO$_3$
is not considered a strongly correlated system, it might be interesting
to investigate the effect of LDA+U in conjunction with the structural
instability.

The authors acknowledge fruitful discussions with D.\ Vanderbilt,  D.\
R.\ Hamann, P.\ Sun, and C.\ J.\ Fennie. This work was supported by ONR
project No.\ N00014--00--01--0261.

\end{document}